\documentstyle[twoside,fleqn,espcrc2]{article}
\newcommand{\pslash}{\mbox{$\not \!p$}}
\newcommand{\qslash}{\mbox{$\not \!q$}}

\newcommand{\AmS}{{\protect\the\textfont2
  A\kern-.1667em\lower.5ex\hbox{M}\kern-.125emS}}
\title{Weak radiative hyperon decays and vector meson dominance}
\author{P. \.{Z}enczykowski\address{Department of Theoretical Physics,
Institute of Nuclear Physics,\\
Radzikowskiego 152, 31-342 Krak\'ow, Poland}}
\begin{document}

\begin{abstract}
We study the question whether the phenomenologically successful
VMD approach to weak radiative hyperon decays
can be made consistent with Hara's theorem and
still yield the pattern of asymmetries exhibited by experimental data.
It appears that an essential ingredient which governs the
pattern of asymmetries is the off-shell behaviour
of the input electromagnetic $1/2^--1/2^+-\gamma $ couplings.
Although this behaviour can be chosen in such a way that
the experimentally observed pattern is obtained, 
and yet Hara's theorem satisfied, at the same time
the approach yields a definite prediction for the size of weak 
meson-nucleon coupling constants. 
Comparison with experiment reveals then another conflict.

\end{abstract}

\maketitle

\section{INTRODUCTION}
Experimental data on weak radiative hyperon decays (WRHD's) present
a challenge to our theoretical understanding.  
The puzzle manifests itself as
a possible conflict between Hara's theorem \cite{Hara} and experiment.

Hara's theorem states that the parity-violating amplitude of the
$\Sigma ^+ \rightarrow p \gamma $ decay should vanish in the limit of SU(3)
flavour symmetry. 
For expected weak breaking of SU(3) symmetry
the parity-violating amplitude in question and, consequently, 
the $\Sigma ^+ \rightarrow p \gamma $ decay asymmetry should be small. 
However, experiment \cite{Foucher} shows that the asymmetry is large:
$\alpha (\Sigma ^+ \rightarrow p \gamma )= -0.72 \pm 0.086 \pm 0.045$.
This large size is even more difficult to explain
when one demands successful simultaneous description of 
data on three related WRHD's, namely
$\Lambda \rightarrow n \gamma $, $\Xi ^0 \rightarrow \Lambda \gamma $, 
and $\Xi ^0 \rightarrow \Sigma ^0 \gamma $.
From the measured size of the WRHD branching ratios
it follows that the single-quark process $s \rightarrow d \gamma $ 
(measured by the $\Xi ^- \rightarrow \Sigma ^- \gamma $
branching ratio) cannot explain the size of the branching ratios
for the $\Sigma ^+$ and neutral hyperon decays.
Consequently, the latter decays must be dominated by (most probably) two-quark  
processes $su \rightarrow ud \gamma $.

Theoretical calculations may be divided into those 
performed totally at quark level 
and those carried out at hadron level.
For a review see ref.\cite{LZ} where recent
theoretical and experimental situation in the field is presented.
It appears that simple quark model calculations 
\cite{KR} violate Hara's theorem.
On the other hand, in pole-model-based hadron-level calculations 
Hara's theorem is usually satisfied by construction \cite{Gavela}.
The only exception here is the hadron-level 
vector-meson dominance (VMD) approach of ref.\cite{Zen}
which admits a pole-model interpretation and yet violates the theorem. 

The WRHD puzzle is deepened by the fact that so far experimental data
seem to agree with the predictions of the VMD model, and not with
those of the (Hara's-theorem-satisfying) standard pole model.
Here, we report on an attempt \cite{Zen98}
to maintain the phenomenological success
of the VMD approach without violating Hara's theorem.

\section{HARA'S THEOREM}
The basic assumptions of Hara's theorem are:
(1) gauge-invariance, (2) CP-conservation, and (3) exact U-spin symmetry.
Since $\Sigma ^+ $ and $p$ differ by the $s \leftrightarrow d$ interchange only,
the $\Sigma ^+$ behaves essentially like a proton. 
Instead of the $\Sigma ^+p\gamma$ coupling,
consider therefore the most general parity-violating $pp\gamma$ coupling 
(see ref.\cite{LZ} for a more
rigorous proof):
\begin{eqnarray}
&\phantom{xxxx}&\overline{\psi} [g_{1}(q^2)
\left(\gamma ^{\mu }-\frac{q^{\mu}\qslash}{q^2} \right)\gamma _5 +\nonumber \\
&& \phantom{xxxx}
g_{2}(q^2)
i\sigma ^{\mu \nu }\gamma _5 q_{\nu } ] \psi \cdot A_{\mu } 
\end{eqnarray}
Since there cannot be a pole at $q^2=0$ (in fact this is assumption (4)), it
follows that $g_1(0)=0$. Furthermore, the $g_2$ term violates CP, and 
therefore we must have $g_2=0$.  Hence, the $pp\gamma $ coupling vanishes
at $q^2=0$. 
Since $\Sigma ^+$ behaves essentially like a proton,
the parity-violating $A(\Sigma ^+ \rightarrow p \gamma )$ 
amplitude must also vanish at $q^2=0$.

\section{MODEL PREDICTIONS \\ AND EXPERIMENT}

Comparison of various model predictions with experiment is given in 
Table 1. Only selected models with the most predictive power are shown.
 
\begin{table*}[hbt]
\setlength{\tabcolsep}{1.1pc}
\newlength{\digitwidth}\settowidth{\digitwidth}{\rm 0}
\catcode`?=\active \def?{\kern\digitwidth}
\caption{ 
Branching ratios (in units of $10^{-3}$) and asymmetries (in italics) 
- comparison of VMD predictions
with those of other models and experiment.  Input values are underlined.
}
\begin{footnotesize}
\begin{tabular*}{\textwidth}{@{}l@{\extracolsep{\fill}}ccccc}
\hline
process \phantom{xxxxxx}
 & experiment & VMD  & quark model  & pole model    & QCD sum rules  \\
       &     & \cite{Zen},\cite{LZ} & \cite{VS} & \cite{Gavela} & \cite{Kha}\\
\hline
&&&&&\\
$\Sigma ^+ \rightarrow p \gamma $ 
  & $\phantom{+}1.23 \pm 0.06$ & $1.26$, $1.4$ & $\underline{1.24}$
                                          & $0.92 ^{+0.26}_{-0.14}$ & $0.8$ \\
  &${\it -0.76 \pm 0.08}$ & ${\it -0.97}$, ${\it -0.95}$ & ${\it -0.56}$ 
  & $ {\it -0.80 ^{+0.32}_{-0.19}}$ & ${\it +1.0}$ \\
$\Lambda \rightarrow n \gamma $ 
  & $\phantom{+}1.63 \pm 0.14$ & $1.0$, $1.7$ & $1.62$ & $0.62$ & $2.1$-$3.1$ \\
  &              & ${\it +0.76}$, ${\it +0.8}$ & ${\it -0.54}$ & ${\it -0.49}$ 
  & ${\it +0.10}$ to ${\it +0.15}$\\
$\Xi ^0 \rightarrow \Lambda \gamma $ 
  & $\phantom{+}1.06 \pm 0.16$ & $0.9$, $1.0$ & $0.50$  & $3.0$   & $1.1$  \\
  & ${\it +0.43 \pm 0.44}$       & ${\it +0.65} $, ${\it +0.80} $ 
  & ${\it +0.68} $ & ${\it -0.78} $ & ${\it +0.9}$ \\
$\Xi ^0 \rightarrow \Sigma ^0 \gamma $ 
  & $\phantom{+}3.56 \pm 0.43$ &   $3.8$,  $4.1$  & $3.30$  & $7.2$   &  \\
  & ${\it +0.60 \pm 0.96^{~\cite{Rambcor}}}$      & ${\it -0.36}$, ${\it -0.45}$ 
  & ${\it -0.94}$ & ${\it -0.96}$ &  \\
$\Xi ^- \rightarrow \Sigma ^- \gamma $
  & $\phantom{+}0.128 \pm 0.023$ & $0.3$, $0.15$ & $\underline{0.23}$ & 
                                                              & $0.1$-$0.2$ \\
  & & ${\it +0.63}$, ${\it +0.7}$ & ${\it -0.6}$ & & ${\it +0.4}$ \\
\hline
\end{tabular*}
\end{footnotesize}
\end{table*}

A closer look at Table 1 reveals that there are two possible patterns of
the signs of asymmetries in the four WRHD's 
($\Sigma ^+ \rightarrow p \gamma $, 
$\Lambda \rightarrow n \gamma $, $\Xi ^0 \rightarrow \Lambda \gamma $, and
$\Xi ^0 \rightarrow \Sigma ^0 \gamma $)
dominated by the $su \rightarrow ud \gamma $ process. 
In Hara's-theorem-satisfying approaches all these asymmetries are of the same
sign (the pole model of ref.\cite{Gavela} predicts the pattern $(-,-,-,-)$).
On the other hand, Hara's-theorem-violating approaches yield the pattern
$(-,+,+,-)$.
Experiment (and, in particular, the sign of 
the $\Xi ^0 \rightarrow \Lambda \gamma $
asymmetry \cite{James}) seems to indicate 
\cite{LZ} that it is the latter alternative that
is realized in Nature. So far the best description of available data
has been provided by the VMD approach \cite{LZ,Zen}.

\setlength{\unitlength}{0.43pt}
\begin{picture}(450,190)
\put(10,0){\begin{picture}(430,170)

\put(0,0){
\begin{picture}(200,160)
\put(100,5){\makebox(0,0){(b1)}}
\put(170,90){\vector(-1,0){25}}
\put(145,90){\line(-1,0){65}}
\multiput(130,65)(0,5){5}{\line(0,1){3}}
\put(135,70){$W$}
\put(80,90){\vector(-1,0){35}}
\put(45,90){\line(-1,0){15}}

\multiput(80,90)(-10,10){5}{\put(0,0) {\oval(10,10)[tr]}
                             \put(0,10){\oval(10,10)[bl]}}
\put(0,130){$\gamma $}

\put(170,65){\vector(-1,0){90}}
\put(80,65){\line(-1,0){50}}
\put(170,40){\vector(-1,0){90}}
\put(80,40){\line(-1,0){50}}
\end{picture}}
\put(230,0){
\begin{picture}(200,160)
\put(100,5){\makebox(0,0){(b2)}}
\put(170,90){\vector(-1,0){15}}
\put(155,90){\line(-1,0){35}}

\multiput(120,90)(-10,10){5}{\put(0,0) {\oval(10,10)[tr]}
                             \put(0,10){\oval(10,10)[bl]}}
\put(40,130){$\gamma $}

\put(170,65){\vector(-1,0){50}}
\put(120,65){\line(-1,0){90}}
\put(170,40){\vector(-1,0){50}}
\put(120,40){\line(-1,0){90}}
\put(120,90){\vector(-1,0){65}}
\put(55,90){\line(-1,0){25}}
\multiput(70,65)(0,5){5}{\line(0,1){3}}
\put(75,70){$W$}
\end{picture}}
\end{picture}
}\end{picture}
\\
\begin{flushleft}
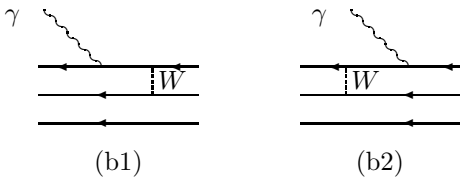
Figure 1. Dominant quark diagrams
\end{flushleft}

The dominant contributions to
the parity-violating WRHD amplitudes
arise from diagrams $(b1)$ and $(b2)$ of Fig.1.  
The difference between Hara's-theorem-violating and Hara's-theorem-satisfying 
models stems from the way in which the amplitudes corresponding to
these two diagrams are calculated and combined.

\section{THE STANDARD AND VMD-BASED POLE MODEL}

\subsection{Standard approach}
In the standard pole model (ref.\cite{Gavela}) the contributions from
diagrams $(b1)$ and $(b2)$ are evaluated at the hadron level as follows 
(consider $\Sigma ^+ \rightarrow p \gamma$ as an example).  The contribution
from intermediate states is approximated by the contribution from excited 
$1/2^-$ baryons $B^*$.  For the $(b1)$ diagram one calculates the weak
transition $\Sigma ^+ \rightarrow N^*$ and the electromagnetic emission 
$N^* \rightarrow p \gamma$ in the quark model.  
The results obtained serve to fix
the coefficients $b$ and $f$ in the relevant hadron-level couplings
$b\overline{}u_{N^*}u_{\Sigma ^+}$ and $f\varepsilon ^*_{\mu} 
\overline{u}_p\sigma ^{\mu \nu} \gamma _5 u_{N^*} q_{\nu}$.
When the contribution from the $(b2)$ ordering is added, the two
contributions together yield the parity-violating amplitude in the form
\begin{equation}
\label{standard}
\left(
\frac{fb}{\Sigma ^+-N^*}-\frac{bf}{p-\Sigma ^*}
\right)
\varepsilon^*_{\mu}\overline{u}_p\sigma ^{\mu \nu} 
\gamma _5 u_{\Sigma ^+} q_{\nu}
\end{equation}
where the two terms stem from diagrams $(b1)$ and $(b2)$ respectively, and
particle names stand for their masses (here $\Sigma ^*$ denotes a 
strange $1/2^-$ baryon).  In the SU(3) limit when $\Sigma^ +=p$ and 
$\Sigma ^* = N^*$, the two contributions cancel ensuring that Hara's theorem
is satisfied.

\subsection{VMD-based approach}
In the VMD approach one calculates first the $\Delta S = 1$
parity-violating coupling of vector mesons ($\rho$,$\omega$,$\phi$).
This is done along the lines of ref.\cite{DDH}.  
The size of all couplings is fixed by symmetry from
the size of the weak nonleptonic hyperon decays.  
The parity-violating couplings of transverse vector mesons to baryons
are identified \cite{DDH} with the hadron-level terms 
$\overline{u}_p\gamma _{\mu} \gamma _5 u_{\Sigma ^+} V^{\mu} $.
In the pole model such terms are obtained using parity-conserving 
baryon-vector-meson couplings $BB^*V$ of the form 
$g\overline{u}_p \gamma _{\mu} \gamma _5 u _{N^*} V^{\mu}$
and lead to the following combination of contributions from
diagrams $(b1)$ and $(b2)$:
\begin{equation}
\label{VM}
\left(
\frac{bg}{\Sigma ^+-N^*}+\frac{gb}{p-\Sigma ^*}
\right)
\overline{u}_p \gamma _{\mu} \gamma _5 u_{\Sigma ^+} V^{\mu}
\end{equation}
In the SU(3) limit ($\Sigma ^+ = p$ and $N^* = \Sigma ^*$) the
contributions from diagrams $(b1)$ and $(b2)$ add rather than subtract.
Consequently, when the VMD prescription $V \rightarrow \frac{e}{g_V}A$
is used to evaluate the parity-violating photon coupling to baryon,
this leads to the appearance of an effective coupling
\begin{equation}
\label{VMD}
g_1(0)\overline{u}_p \gamma _{\mu} \gamma _5 u_{\Sigma ^+} A^{\mu}
\end{equation}
with a nonvanishing $g_1(0)$. This violates the assumptions upon which 
Hara's theorem is based.
Expression (\ref{VMD}) may be considered a part of a gauge-invariant
coupling $g_1(0)\overline{u}_p (\gamma _{\mu} 
-\qslash q^{\mu}/q^2)\gamma _5 u_{\Sigma ^+} A^{\mu}$
(for transverse photons $q \cdot A=0$).
The presence of the pole at $q^2=0$ may be
worrying, however, as no massless hadrons exist.

\subsection{Comparison of two approaches}
Subtraction, Eq.(\ref{standard}), (respectively addition, Eq.(\ref{VM})) 
of pole expressions corresponding to diagrams $(b1)$ and $(b2)$ leads to 
asymmetry patterns $(-,-,-,-)$ (respectively $(-,+,+,-)$).
Thus, the pattern $(-,+,+,-)$ seems to signify the violation of Hara's
theorem.

\section{SATISFYING HARA'S THEOREM \\ WITH $(-,+,+,-)$ PATTERN OF
ASYMMETRIES}

So far the experiment seems to confirm the $(-,+,+,-)$
pattern of asymmetries.  In detailed models (VMD, quark model) this
pattern signifies violation of Hara's theorem.
The origin of this violation is, however, slightly different in the two
models \cite{Acta}.  The quark model calculations \cite{KR} directly
violate the theorem.
VMD may be considered more phenomenological.  Thus,
the question emerges whether the asymmetry pattern $(-,+,+,-)$ obtained
in the VMD approach
could be maintained and yet Hara's theorem satisfied. This question 
has been discussed recently in ref.\cite{Zen98}.

Let us reconsider the problem of the most general parity-conserving
gauge-invariant coupling of a real photon to $1/2^+ - 1/2^-$ baryonic current.
Although it seems that there is only one such coupling that does not involve
the pole at $q^2=0$, ie.
\begin{equation}
\label{j1}
\overline{u}_{B_k}i\sigma _{\mu \nu} \gamma _5 u_{B^*_l} q^{\mu} A^{\nu}~,
\end{equation}
in fact one also has to consider
\begin{eqnarray}
\label{eq:basicidentity}
\lefteqn{
h(-i) 
(p_k+p_l)_{\lambda }q_{\nu }
\epsilon ^{\lambda \mu \nu \rho}\overline{u}_{B_k}\gamma _{\rho } u_{B^*_l} 
A_{\mu}\equiv}
\\
&&h\overline{u}_{B_k}(q^2\gamma ^{\mu }-q^{\mu }\qslash)\gamma _5 u_{B^*_l}
A_{\mu}+ \nonumber \\
&&
h\overline{u}_{B_k}(\pslash_k
i\sigma ^{\mu \nu }\gamma _5 q_{\nu }-
i\sigma ^{\mu \nu }\gamma _5 q_{\nu }\pslash_l) u_{B^*_l}A_{\mu}~.
\nonumber
\end{eqnarray}
For $q^2=0$, $q \cdot A=0$ and on-mass-shell baryons,
expression (\ref{eq:basicidentity}) 
reduces to that of (\ref{j1}) (barring some mass factors).
However, such a reduction cannot be effected in pole model calculations 
of WRHD's since $B^*$ is {\em not} on its mass shell.
One has to keep expression (\ref{eq:basicidentity}) in the calculations
and only at the end use the fact that external baryons are on their mass shells.
Calculation \cite{Zen98} 
gives then for the parity-violating amplitude in the weak
$i \rightarrow f \gamma $ decay the expression
\begin{eqnarray}
\label{alternative}
(m_i-m_f)\left(
\frac{hb}{m_i-m_{B^*}}+\frac{bh}{m_f-m_{B^*}}
\right)\times &&\nonumber \\
\overline{u}_f i\sigma ^{\mu \nu} q _{\nu} \gamma _5 u_i A_{\mu}&&
\end{eqnarray}
From the form of Eq.(\ref{alternative})
we see that one can maintain the pattern $(-,+,+,-)$ (in which contributions
from $(b1)$ and $(b2)$ add rather than subtract) and yet have Hara's theorem
 satisfied (this is ensured by the overall factor of $m_i-m_f$).  In this case
 all (and not just $\Sigma ^+ \rightarrow p \gamma$)
 parity-violating WRHD amplitudes vanish in the SU(3) limit.
 In comparison with the Hara's-theorem-violating approach of refs.\cite{Zen,LZ},
 the difference consists in the presence of an additional 
 $(m_i-m_f)^2$ factor in the amplitudes.
 Clearly, the existence of such an overall mass factor cannot be
 experimentally verified since we cannot
 move the mass of $\Sigma ^+$ to come closer to that of
 the proton.
 However, we can look at the $\Delta S = 0$ processes, ie. at the weak coupling 
 of mesons to nucleons (such as $\rho N N$ etc.).  If the relative factor of
 $(m_i-m_f)^2$ is there, the weak $\rho ^+ p n$ coupling would be scaled down
 with respect to that estimated by symmetry from nonleptonic hyperon decays
 by a factor of
 \begin{equation}
 \left(
 \frac{m_n-m_p}{m_{\Sigma}-m_p}
 \right)^2
 \approx 10^{-4}
 \end{equation}
Thus, the resulting weak $\rho N N$ coupling would be totally negligible.
The data indicate, however, that the scale of the $\Delta S =0$ coupling
is the same as that of the $\Delta S =1$ couplings \cite{AH85}.
Thus, there is no such mass factor and the observation of the
$(-,+,+,-)$ pattern of asymmetries in WRHD's will signify violation of
Hara's theorem.

\section{CONCLUSIONS}
Theoretical VMD predictions for the asymmetries of the
$\Xi ^0 \rightarrow \Lambda \gamma$ and 
$\Xi ^0 \rightarrow \Sigma ^0 \gamma$ decays are very
solid (errors cannot exceed 0.2) \cite{LZ}. 
There are experimental indications \cite{Ramberg} that in the latter case
the asymmetry is indeed moderately negative, in agreement with VMD predictions.
Although the VMD approach can be made consistent with Hara's theorem
and still yield positive asymmetry parameter in the 
$\Xi ^0 \rightarrow \Lambda \gamma$ decay, this requires
vanishingly small weak meson-nucleon couplings, in gross disagreement
with experiment.
It is therefore very important to measure the 
$\Xi ^0 \rightarrow \Lambda \gamma$ asymmetry precisely.
This asymmetry is {\it significantly} positive (negative) for all models
violating (satisfying) Hara's theorem.  If the KTeV experiment 
\cite{Ramberg,Newprop99} finds here
a negative value, perhaps one can 
concoct a model which satisfies Hara's theorem and describes the data.
If, on the other hand, the KTeV experiment reports a positive value,
Hara's theorem is violated in Nature.  
Questions concerning the meaning of this violation may then be more
legitimately asked.



\begin{thebibliography}{99}
\bibitem{Rambcor} The published experimental number is corrected here 
for a missing factor of 3, (E. Ramberg, these proceedings).
\end{thebibliography}
\end{document}